\documentstyle[12pt]{article}

\begin{document}
\begin{titlepage}
\begin{center}

November 27, 2001     \hfill    LBNL-49171 \\

\vskip .5in

{\large \bf Parity Doubling, Zitterbewegung, and Rest Mass for Spin 1 }

\footnote{This work is supported in part by the Director, Office of Science, 
Office of High Energy and Nuclear Physics, Division of High Energy Physics, 
of the U.S. Department of Energy under Contract DE-AC03-76SF00098}

\vskip .50in
Geoffrey F. Chew and Henry P. Stapp\\
{\em Lawrence Berkeley National Laboratory\\
      University of California\\
    Berkeley, California 94720}
\end{center}

\vskip .5in

\begin{abstract}
A 6-component ``wave function" (not field, but S-matrix interpretable) for 
a massive spin-1 particle parallels the Dirac ``chirality-doubled" 
4-component wave function for a spin-1/2 particle, by pairing two wave 
functions for same spin but opposite ``handedness". The correlated 
``opposite-parity" pair of complex 3-vectors defines a fluctuating 
spin-correlated lightlike ``internal velocity" as well as an independent 
``external rapidity". Extension from fermions to vector bosons of the 
velocity-fluctuation (``zitterbewegung") interpretation of rest mass weakens 
theoretical motivation for elementary scalar bosons.  

\end{abstract}

\end{titlepage}

\newpage
\renewcommand{\thepage}{\arabic{page}}
\setcounter{page}{1}

\noindent {\bf Introduction}

	The standard model accommodates massive spin-1 particles, with zero as well as $\pm 1$ helicity, by enlarging the fermion, vector-boson family of elementary particles to include scalar bosons$.^{(1)}$ The present paper employs a 6-dimensional representation of spin 1---imitating the wave-function doubling that characterizes Dirac 4-spinors (a doubling that allows representation not only of parity but of charge conjugation)---to define a Hilbert space that accommodates (charged or neutral) spin 1 with nonvanishing rest mass even though based on elementary ``lightlike" (zero-rest-mass) vectors (no scalars). Rest mass, for vector bosons as well as for fermions, is attributed to ``zitterbewegung".  

Dirac's 4-component chirally-doubled-spinor wave function for a spin-1/2 particle recognizes a momentum-independent ``internal" significance for velocity that parallels or antiparallels spin (not momentum) in its direction. The unique internal-velocity magnitude is that of light (Dirac's ``velocity operator" is 
$\gamma_5 \vec{\sigma}).^{(2)}$ The Dirac wave function for nonzero rest mass manifests a fluctuation, at fixed spin (and momentum), of a 2-valued internal-velocity direction; only with zero rest mass does internal velocity fail to fluctuate. (The expectation value of Dirac's internal-velocity operator is equal to the ratio of momentum to energy---i.e., to ``external velocity".) Dirac's association of rest mass with lightlike internal-velocity fluctuation---``zitterbewegung"---subsequently became obscured by continuous-field theory but accords with a dynamics, based on lightlike discrete-step Feynman paths, that is under investigation by one of the authors. 

We here call attention to a spin-correlated directional fluctuation of lightlike internal velocity within the wave function for a spin-1 particle with nonzero rest mass. A correlated pair of ``opposite-parity" complex 3-vector wave functions---a 6-component wave function---depicts the ``source" of rest mass in a sense similar to that of Dirac. Although our meaning for internal velocity is unaccompanied by any vector-boson Hamiltonian, wave-function zitterbewegung propagation is prescribable, even in absence of either single-particle Hamiltonian or field-theoretic Lagrangian, by discrete Feynman paths. Zitterbewegung ``source" of rest mass for vector-bosons as well as fermions weakens theoretical motivation for elementary scalar bosons.

Orthogonal to the enterprise here, the much earlier Duffin-Kemmer-Petiau (DKP) massive spin-1 representation$^{(3)}$ ignored the concept of ``elementary zero-rest-mass constituents" while imitating Dirac by demanding a Hamiltonian. The DKP representation is founded on the real (1/2, 1/2) 4-vector representation of the Lorentz group and requires a 10-component wave function in the simplest form found. DKP exhibits the familiar spin 0-spin 1 association through spin-space representations that differ from the (0, 1) + (1, 0) employed in the present paper.

Our proposed vector-boson wave function reflects a correspondence between a real second-rank antisymmetric Lorentz tensor and a pairing (direct sum) of complex 3-vectors with opposite behaviors under Lorentz boosts. The complex 6-vector allows intertwined Lorentz-group representation of particle spin and particle motion. There is no need to postulate an equation of motion (e.g., Klein-Gordon equation) that requires meaning for continuous time. Time is allowed to have a discrete meaning such as, for example, the ``in-out" S-matrix sense.

The complex 3-vector is a device used sometimes in classical electromagnetic theory but is there merely a convenience. Complexity, on the other hand, is essential to our wave-function representation of particle spin and allows internal-velocity reversal by parity inversion to associate for spin 1 (as it does for spin 1/2) with hermitian-conjugate inverse of Lorentz-group representation (similar to complex conjugation). We shall find also contributing to internal-velocity fluctuation, for vector bosons although not for fermions, a discrete lightlike-velocity reversal corresponding to $180^{\circ}$ rotation. 

The Dirac 4-spinor admits representation of the complex Lorentz group-a 12-parameter group that includes 4-vector inversion$^{(4)}$---and thereby indirectly represents charge conjugation through CPT. Although not here pursued, the 6-vector proposed in what follows similarly admits CPT representation through the complex Lorentz group. 
	
A Lorentz-invariant constraint on a general complex ``handed" 3-vector---a (0,1) or (1,0) representation of the Lorentz group---reduces by 2 the number of independent real parameters and achieves, for what we here call a ``rotor", 4-parameter lightlike (zero-rest-mass) single-helicity status paralleling that of a 2-spinor---i.e., paralleling that of a (0, 1/2) or (1/2, 0) representation. (The terms ``rotor" and ``elementary handed vector" are to be understood as synonymous.) The most general handed complex 3-vector is a superposition of 2 rotors with common helicity but opposite 3-velocity. Furthermore, as with spinors, parity doubling allows in a special Lorentz frame a pairing of opposite-velocity rotors with same spin and thus opposite helicity---``left-handed" and ``right-handed". Dirac associated the pairing (direct sum) of opposite-velocity 2-spinors with a spin-1/2 particle of nonzero rest mass. We here shall represent a massive spin-1 particle through opposite-veloci!
ty rotor pairing and superposition. 
	 
The usual particle concept defines helicity through an ``external" velocity---equal to the ratio of momentum to energy---and accords no meaning to internal velocity. Pairing and superposition of opposite internal lightlike velocities is nevertheless amenable to propagation via classical Feynman paths that exhibit the Chan-Paton structure (5) underpinning SU(N) ``internal symmetry" (e.g., isospin or color). For massive vector bosons, the present paper not only invokes Dirac-like pairing of opposite internal helicities but finds zero ``external helicity", as well as helicity $\pm 1$, to be represented through (same-handed) opposite-velocity rotor superpositions.

The proposal here (attributing lightlike velocity and single nonvanishing helicity to any ``elementary particle") offers no clue as to why internal-velocity fluctuation fails to occur for photons (whose internal and external velocities coincide). A Feynman-path zitterbewegung propagation mechanism  (not involving elementary scalar bosons), with potential for explaining rest-mass magnitudes, will be put forward elsewhere. The mechanism recognizes not only special status for the photon (related to its electric-charge coupling) but the here-ignored although general helicity-reversing CPT matching of particle with antiparticle. Any (zero-rest-mass) elementary particle is matched by a different elementary particle---its CPT ``antiparticle"---characterized by opposite helicity as well as by opposite ``charges". The particle ``anti" to a photon is a photon of opposite helicity although the same zero value for all charges. 

 Despite our recognition of CPT, no meaning is anticipated for a local rotor-field. CPT invariance and the connection between spin and statistics do not require local field theory$.^{(6)}$ Our elsewhere-described Feynman-path mechanism for zitterbewegung rests on lightlike discrete path steps that dovetail with discreteness of wave-function time. There is no continuous-time evolution generated by a Hamiltonian (or meaning for Lagrangian). Fields continuous in time (with time derivatives) need be meaningful only in some large-scale approximation.\\  

\noindent {\bf Definition of ``Rotor" as a Lightlike Complex 3-Vector}

	A general complex 3-vector $\vec{W}$ has 6 real parameters. It is pedagogically helpful to distinguish right and left-handed complex 3-vectors, not only as being (0,1) or (1,0) representations of the Lorentz group, but by a notational device employing the familiar real 3-vector symbols $\vec{E}$ and $\vec{H}$. In either right or left-handed {\bf (ket)} 3-vector, the symbol $\vec{H}$ will designate the real part, whereas the symbol $-\vec{E} (+\vec{E})$ designates the right-handed (left-handed) imaginary part. That is
$$
\vec{W}_r = \vec{H}_r - i\vec{E}_r ,  \eqno (1.a)
$$
$$	   	                           
\vec{W}_l  = \vec{H}_l + i\vec{E}_l  . \eqno (1.b)
$$                                                                                                    
(Wherever the Dirac bra-ket notation is not employed in this paper, the symbols
$\vec{W}_r$ and $\vec{W}_l$ are to be understood as ket vectors.) Lorentz boost of either type of complex 3-vector accords with that of electric and magnetic fields designated by these symbols. Specifically, writing a (real) 3-vector boost parameter $\vec{b}$ as $\beta \vec{u}$ , where $\vec{u}$ is a unit 3-vector, the boost behavior of our $\vec{H}$ and $\vec{E}$ symbols (for both $r$ and $l$ and for both bra and ket) is
$$
\vec{H}_{\vec{b}} = \cosh\beta \, \vec{H}_0 + ( 1- \cosh\beta ) (\vec{u}\cdot \vec{H}_0 )\vec{u} + \sinh \beta (\vec{u}\times\vec{E}_0 ) \eqno (2.a)
$$
$$
\vec{E}_{\vec{b}} = \cosh\beta \, \vec{E}_0 + ( 1- \cosh\beta ) (\vec{u}\cdot \vec{E}_0 ) \vec{u} - \sinh\beta (\vec{u}\times \vec{H}_0) . \eqno (2.b)
$$           

The difference, according to (1), of $\vec E$ symbol meaning in right and left-handed complex 3-vectors represents ``opposite-parity" behavior of these two categories under boosts---the boost-direction parameter $\vec{u}$ reversing sign. (An alternative equivalent formalism may be based on bilinear products of (0,1/2) spinors and of (1/2, 0) spinors.)  

In contrast, transformation of any complex 3-vector under rotations is generated by the standard purely-imaginary $3\times 3$ hermitian traceless spin-1 matrices$^{(7)}$ that here we denote by the 3-vector symbol $\vec{S}$. The symbols $\vec{E}$ and $\vec{H}$ transform in parallel (without mixing with each other) under rotations, whose 3-dimensional representation is real. 

In terms of $\vec{S}$, the right and left-handed $3\times 3$ positive-definite hermitian unimodular matrix representations of a (ket) boost by $\vec{b}$---equivalent to (2)---are 
$$
B_{\pm}(\vec{b}\,) = \exp (\pm \vec{S}\cdot \vec{b}\,) = I + (\cosh\beta -1)(\vec{S} \cdot \vec{u}\,)^2 \pm \vec{S} \cdot \vec{u} \sinh \beta . \eqno (3)
$$   
The relation $(\vec{S} \cdot \vec{u}\,)^3 = \vec{S}\cdot\vec{u}$ has been used 
to expand the exponential. Notice that $B_+(\vec{b}\,)\,B_-(\vec{b}\,)  = I$. The righthanded (lefthanded) representation carries the positive (negative) subscript in (3). What we here are calling ``parity inversion" corresponds to complex conjugation of our Lorentz-group representation.

A general complex 3-vector $\vec{W}$ (either $r$ or $l$) has 6 real parameters. An $r$ or $l$  ``rotor" is a 4-parameter complex 3-vector satisfying the Lorentz-invariant constraint,
$$
	\vec{W} \cdot \vec{W} = 0 ,	   \eqno (4)
$$
or, equivalently,
$$
(Re\:\vec{W})^2-(Im\:\vec{W})^2=0,\:\:\: 2 Re\:\vec{W}\cdot Im\:\vec{W}=0. \eqno (4')        $$                      
For either left or right-handed rotors, the foregoing implies
$$	
(\vec{H})^2 = (\vec{E})^2,\:\:\:  \vec{H} \cdot \vec{E} = 0. \eqno (4'')            $$
                                   
Paralleling a 2-spinor, the 4 real parameters of a rotor correspond to 3 Euler angles together with complex-vector magnitude. Two of the angles prescribe both the spin direction (next paragraph) and the lightlike 3-velocity,  
$$
 \vec{w} = 2 (\vec{E}\times\vec{H})/(\vec{E}^2+ \vec{H}^2) ,  \eqno (5)           
$$
of an ``elementary vector boson" with zero rest mass; the remaining angle 
controls the rotor phase. In the right-handed (left-handed) category, velocity direction parallels (antiparallels) spin direction. Rotation about the spin direction changes the rotor phase by an increment equal to the angle of rotation. The rotor magnitude varies with the Lorentz frame (as also does the magnitude of a 2-spinor); for example, boosting in the direction of $\vec{w}$ increases the magnitude of both $\vec{E}$ and $\vec{H}$ by the factor $\exp \beta$. 

	Although any superposition of same-handed rotors that share the same (real) 3-velocity $\vec{w}$---i.e., of rotors related simply by multiplicative factors---is another rotor of this same handedness and velocity, a superposition of 2 same-handed different-velocity rotors is not a rotor. In fact, any right- or left-handed complex 3-vector is a superposition of 2 rotors of the same handedness and opposite velocities. For nonparallel $\vec{E}$ and $\vec{H}$, these two velocities are perpendicular to the plane containing $\vec{E}$ and $\vec{H}$.     

	An illuminating alternative way to understand the meaning of ``rotor" is through the notion of ``infinite boost" applied to a general handed complex 3-vector $\vec{W}_0$. It may be verified from (2) or (3) that, so long as $\vec{E}_0$, $\vec{H}_0$, and boost-direction $\vec{u}$ are not all collinear, the limit of $(\exp -\beta) \: \vec{W}_{\beta \vec{u}}$ as $\beta \rightarrow \infty$ is a (correspondingly-handed) rotor with 3-velocity $\vec{u}$. The foregoing infinite-boost limit confirms the consistency of associating (5) with rotor 3-velocity and, as noted below, allows a unified description of zero and nonzero rest mass. 
  
The $3\times 3$ hermitian matrices $\vec{S}$, that generate spin-1 rotations, have for a general handed complex 3-vector the (easily-calculated) expectation values$$   
<\vec{W}|\vec{S}|\vec{W} >=2 Re\,\vec{W}\times Im\,\vec{W} = \pm 2(\vec{E}\times\vec{H}) ,  \eqno (6)
$$                
with                      
$$   
< \vec{W} |\vec{W}> = \vec{E}^2+\vec{H}^2 .  \eqno (7)
$$
The positive (negative) sign in (6) attaches to right-handed (left-handed) complex 3-vectors. The rotor characteristic $(4'')$ renders (7) equal to the modulus of (6) and thereby not only makes $\vec{w}$ a unit vector but implies any right-handed rotor to be an eigenvector of $\vec{S}\cdot \vec{w}$ with eigenvalue $+1$, while any left-handed rotor belongs to the eigenvalue $-1$. Unit velocity is ``light velocity". Any rotor is a 3-component quantum state of an elementary vector boson with zero rest mass and well-defined helicity.\\
 
\noindent {\bf Spin-1 Parity-Doubled Spin-Rapidity Wave Function for Nonzero Rest Mass}

Equation (6) indicates zero average spin for any (handed) complex 3-vector whose real and imaginary parts are parallel. Such states in fact are eigenvectors with zero eigenvalue of the spin component in the foregoing direction but, in contrast to rotors, no single 4-velocity associates with such a state. How can we represent the state of a massive spin-1 particle moving (at sub-light velocity) with (external) helicity that may be either zero or ±1?   
 
Imitating Dirac, we mutually associate a right-left pair of complex 3-vectors through a rapidity parameter to build a single-particle Hilbert space of complex 6-vectors. Rapidity is equivalent to a unit-norm positive-timelike 4-vector and thus transforms as a real (1/2, 1/2) representation of the Lorentz group. From a rapidity basis, passage to a momentum or coordinate basis is straightforward. A rapidity $\vec{b}$ (in some specified Lorentz frame) corresponds to an external sub-light 3-velocity $\vec{u} \tanh \beta$. As was true for Dirac, there is equality at zero rapidity of the paired left and right-handed complex 3-vectors while for any rapidity the left and right-handed wave-function components relate to a common normalized (``unhanded") complex 3-vector by opposite boost transformations prescribed below in a Lorentz-frame independent manner. The particle rapidity determines the relation between right and left components (and vice-versa). The common normalized complex !
3-vector may for zero-rapidity be described as ``the rest-frame spin wave function".

	In a rapidity basis as opposed to a momentum basis, rest mass fails to be specified; dimensionality is lacking. Rest-mass is a characteristic of wave-function propagation as time advances but not of the wave function itself in rapidity space; rapidity is dimensionless. Multiplication of external 4-velocity $(\cosh \beta, \vec{u} \sinh \beta)$ by rest mass achieves a 4-momentum whose 3 ``space components" may be regarded as equivalent to rapidity. Passage between 3-momentum and (spatial) coordinate bases by Fourier transformation is then standard. (Opposite dimensionality of coordinate and momentum relates to the rest-mass setting of a length scale---the ``Compton wavelength".)

 To the extent that its rapidity is always infinite the wave function of a vector boson whose rest mass vanishes might seem not to admit a rapidity basis. But arbitrary smallness of nonvanishing rest mass is compensatable at any fixed momentum by arbitrary largeness of finite rapidity. The 4-parameter right or left-handed rotor emerging in a fixed-momentum infinite-rapidity limit (related to the above-described infinite-boost limit) as rest mass approaches zero parallels the 2-spinor representation of a massless elementary fermion.  

 	For our 6-component parity-doubled wave function in Lorentz-frame $\Sigma$, we use the notation
$$ 
\Psi^{\Sigma}(\vec{b}\,) = [\vec{W}^{\Sigma}_l(\vec{b}\,),\vec{W}^{\Sigma}_r (\vec{b}\,)] ,   \eqno (8)
$$             
where, paralleling the representation of the Dirac wave function given in Reference (6),
$$ 
\vec{W}^{\Sigma}_l(\vec{b}\,)=B_-(\vec{b}\,)\vec{W}^{\Sigma}_0(\vec{b}\,),\eqno(9.a)
$$ 
and
$$
\vec{W}^{\Sigma}_r(\vec{b}\,)=B_+(\vec{b}\,)\vec{W}^{\Sigma}_0(\vec{b}\,).\eqno(9.b) $$ 

The pair of handed complex 3-vectors $\vec{W}^{\Sigma}_r (\vec{b}\,)$  and $\vec{W}^{\Sigma}_l (\vec{b}\,)$ are, respectively, results of the right and left-handed boosts with parameter $\vec{b}$ of an ``unhanded" complex 3-vector rapidity-function $\vec{W}^{\Sigma}_0 (\vec{b}\,)$ whose frame-independent norm is prescribed below. Note that, because $B_-(\vec{b}\,)$ is the inverse of $B_+(\vec{b}\,)$, $\vec{W}^{\Sigma}_l (\vec{b}\,) = B_-(2\vec{b}\,) \vec{W}^{\Sigma}_r (\vec{b}\,)$.
 
Invariance of the relations (9) under change of Lorentz frame is verifiable by straightforward (although nontrivial) calculation. Within the (elementary-vector-boson) rotor decomposition of $\vec{W}^{\Sigma}_0 (\vec{b})$, each ``constituent rotor" appears for any rapidity in both right and left components of $\Psi^{\Sigma}(\vec{b})$ although with a rapidity-dependent magnitude that differs between right and left. A constituent rotor plays a role paralleling that of a 2-spinor in Dirac's 4-component wave function.

  Equality of left and right-handed $\Psi^{\Sigma} (\vec{b}\,)$ components at zero $\vec{b}$ means, with rotor decomposition, light velocities of opposite directions for matched (same spin) $l$ and $r$ constituent rotors. Averaging the two lightlike 4-velocities, with equal time components, of any matched pair then yields at zero $\vec{b}$ for this pair a positive purely-timelike 4-velocity (i.e., zero 3-velocity). In this sense, just as for Dirac, the ``average internal 3-velocity" vanishes when rapidity vanishes. Boosting the ``average internal 4-velocity" from the rest frame defines the average internal rapidity to be $\vec{b}$. Right-left internal-velocity difference we interpret (with Dirac) as rapidity-controlled fluctuation. For massive vector bosons there is additionally a spin-controlled same-handed elementary-vector velocity fluctuation. 

	  Further imitation of Dirac recognizes the 6-dimensional internal Hilbert space to be factorizable into the direct product of a 3-dimensional spin space and a 2-dimensional ``parity space". ($2\times 2$ parity matrices commute with $3\times 3$ spin matrices although not generally with each other.) It is convenient to define a diagonal parity matrix $\gamma_5$ with the right-hand eigenvalue +1 and the left-hand eigenvalue -1. The $6\times 6$ ``Poynting 3-vector" matrix $\gamma_5 \vec{S}$ [see Formula (6)] then joins the $6\times 6$ ``energy-density" unit-matrix [see Formula (7)] and the $6\times 6$ representation of spin (the product of $\vec{S}$ with the unit parity matrix). Representation of left-right transitions in the zitterbewegung generation of rest mass involves the off-diagonal real-symmetric self-inverse ``parity-inversion" matrix $\gamma_0$ that anticommutes with $\gamma_5$. The ``spin-space expectation"
$$ 
N^{\Sigma}(\vec{b}\,) =<\Psi^{\Sigma}(\vec{b}\,)|\gamma_0 |\Psi^{\Sigma}(\vec{b}\,)>  =  2 <\vec{W}^{\Sigma}_0 (\vec{b}\,)|\vec{W}^{\Sigma}_0 (\vec{b})\,> \eqno (10)    
$$
is a positive-definite Lorentz-scalar function of $\vec{b}$, i.e., its value in a transformed frame $\Sigma'$ at the transformed point $\vec{b}'$ is independent of $\Sigma'$. The integral of (10) over rapidity space, with Lorentz-invariant volume element $d\vec{b}$, provides a frame-independent norm for  $|\Psi >$.

	Defining the average rapidity for a particle in the state $\Psi$ as measured in Frame $\Sigma$ by
$$	
\vec{b}^{\,\Sigma}\equiv \int d\vec{b}\:N^{\Sigma}(\vec{b}\,)\:\vec{b},\eqno (11)
$$
one may define an average ``energy-flux 3-velocity" (``momentum density" divided by ``energy density") for the particle by the spin-space expectation quotient
$$
<\Psi^{\Sigma}(\vec{b}^{\,\Sigma})|\gamma_5 \vec{S}|\Psi^{\Sigma}(\vec{b}^{\,\Sigma}) > /<\Psi^{\Sigma}(\vec{b}^{\,\Sigma})|\Psi^{\Sigma}(\vec{b}^{\,\Sigma})>.\eqno (12)$$                                          
The foregoing quotient vanishes for zero average rapidity and approaches light velocity with direction $\vec{u}^{\,\Sigma}$ in the limit as $\beta^{\Sigma}\rightarrow \infty$, but defines a 3-vector that under boosts transforms differently from average external velocity (except in the infinite average-rapidity limit)---attaching to a symmetric second-rank (``energy-momentum-stress") tensor rather than to a positive-timelike 4-vector.  For Dirac, the coordinate-basis spin-space expectation of whose $4\times 4$ unit matrix transforms like probability density (or charge density) rather than energy density, the analog of (12) equals average external velocity.\\

\noindent {\bf Conclusion}
 
	The 6-component wave function $\Psi^{\Sigma}(\vec{b}\,)$, in a sense resembling that of Dirac's 4-component wave function, decomposes into ``elementary constituents" that provide meaning for fluctuating lightlike ``internal velocity" correlated with spin. The rapidity basis for $\Psi$ provides a meaning for massive-particle external velocity that is independent of spin. When rapidity direction fails to parallel or to antiparallel a (nonfluctuating) spin direction, the wave function has zero-helicity components even though any $\Psi$ is decomposable into lightlike (helicity $\pm 1$, elementary-vector) rotors.
            Despite important differences between Dirac's 4-component spin-1/2 wave function and the 6-component spin-1 wave function proposed here, similarities encourage the attitude that zitterbewegung generally underpins rest mass. A non-Lagrangian dynamics, based on lightlike discrete-step (action-carrying) Feynman paths that admit velocity reversal, may encounter no need for elementary scalar bosons in the generation of either vector-boson or fermion rest mass.\\
                                                                                                                                                            
\noindent {\bf Acknowledgements}

Advice from J. D. Jackson, M. Chanowitz and B. Zumino, unpublished lecture notes by E. Wichmann and discussions with J. Finkelstein have contributed to this paper. \\

\noindent {\bf References}\\
1. S. Weinberg, {\it The Quantum Theory of Fields}, Vol. II, Cambridge University Press, New York (1996).\\
2. P. A. M. Dirac, {\it Quantum Mechanics}, third edition, p. 260, Oxford University Press, New York (1947).\\
3. H. Umezawa, {\it Quantum Field Theory}, p. 85, North Holland, Amsterdam (1956).\\
4. R. F. Streater and A. S. Wightman, {\it PCT, Spin and Statistics and All That}, Benjamin and Co., New York (1964).\\
5. J. E. Paton and Chan, H. M., Nucl. Phys. B10, 516 (1969).\\
6. H. P. Stapp, Phys. Rev. 125, 2139 (1962).\\
7. L. Schiff, {\it Quantum Mechanics}, third edition, p. 198, McGraw-Hill, New York (1968).

\end{document}